# Compensation of loss in propagating surface plasmon polariton by gain in adjacent dielectric medium

M. A. Noginov[1*], V. A. Podolskiy[2], G. Zhu[1], M. Mayy[1], M. Bahoura[1], J. A. Adegoke[1], B. A. Ritzo[1], K. Reynolds[1]

[1] *Center for Materials Research, Norfolk State University, Norfolk, VA 23504*

[*] *mnoginov@nsu.edu*

[2] *Department of Physics, Oregon State University, Corvallis, OR 97331-6507*

**Abstract:** We report the suppression of loss of surface plasmon polariton propagating at the interface between silver film and optically pumped polymer with dye. Large magnitude of the effect enables a variety of applications of 'active' nanoplasmonics. The experimental study is accompanied by the development of the analytical description of the phenomenon and the solution of the controversy regarding the direction of the wavevector of a wave with a strong evanescent component in an active medium.

Surface plasmon polaritons (SPPs) – special type of electromagnetic waves coupled to electron density oscillations – allow nanoscale confinement of electromagnetic radiation [1]. SPPs are broadly used in photonic and optoelectronic devices [1-7], including waveguides, couplers, splitters, add/drop filters, and quantum cascade lasers. SPP is also the enabling mechanism for a number of negative refractive index materials (NIMs) [8-12].

Many applications of SPPs suffer from damping caused by absorption in metals. Over the years, several proposals to compensate loss by incorporating active (gain) media into plasmonic systems have been made. Theoretically, field-matching approach was employed to calculate the reflectivity at surface plasmon excitation [13]; the authors of [14] proposed that the optical gain in a dielectric medium can elongate the SPP's propagation length; gain-assisted excitation of resonant SPPs was predicted in [15]; SPP propagation in active waveguides was studied in [16]; and the group velocity modulation of SPPs in nano-waveguides was discussed in [17]. Excitation of localized plasmon fields in active nanosystems using *s*urface *p*lasmon *a*mplification by *s*timulated *e*mission of *r*adiation (SPASER) was proposed in [18]. Experimentally, the possibility to influence SPPs by optical gain was demonstrated in Ref. [19], where the effect was as small as 0.001%.

Here we report conquering the loss of *propagating* SPPs at the interface between silver film and optically pumped polymer with dye. The achieved value of gain, 420 cm$^{-1}$, is sufficient to fully compensate the *intrinsic* SPP loss in high-quality silver films. This, together with the compensation of loss in *localized* surface plasmons, predicted in [20] and recently demonstrated in [21], enables practical applications of a broad range of low-loss and no-loss photonic metamaterials.

The experimental attenuated total reflection (ATR) setup consisted of a glass prism with the real dielectric permittivity $\varepsilon_0 = n_0^2$, a layer of metal with the complex dielectric constant $\varepsilon_1$ and thickness $d_1$, and a layer of dielectric medium characterized by the permittivity $\varepsilon_2$, Fig. 1a.

The wave vector of the SPP propagating at the boundary between media 1 and 2, is given by [1]

$$k_x^0 = \frac{\omega}{c} \sqrt{\frac{\varepsilon_1 \varepsilon_2}{\varepsilon_1 + \varepsilon_2}}, \qquad (1)$$

where ω is the oscillation frequency and $c$ is the speed of light. SPP can be excited by a $p$ polarized light falling on the metallic film at the critical angle $\theta_0$, such that the projection of the wave vector of the light wave to the axis $x$,

$$k_x(\theta) = (\omega/c) n_0 \sin\theta_0, \tag{2}$$

is equal to $\mathrm{Re}(k_x^0)$. At this resonant condition, the energy of incident light is transferred to the SPP, yielding a minimum (dip) in the angular dependence of the reflectivity $R(\theta)$ [1]:

$$R(\theta) = \left| \frac{r_{01} + r_{12} \exp(2ik_{z1}d_1)}{1 + r_{01}r_{12} \exp(2ik_{z1}d_1)} \right|^2, \tag{3}$$

where $r_{ik} = (k_{zi}\varepsilon_k - k_{zk}\varepsilon_i)/(k_{zi}\varepsilon_k + k_{zk}\varepsilon_i)$ and

$$k_{zi} = \pm\sqrt{\varepsilon_i\left(\frac{\omega}{c}\right)^2 - k_x(\theta)^2}, \quad i = 0,1,2. \tag{4}$$

The parameter $k_z c/\omega$ defines the field distribution along the $z$ direction. Its real part can be associated with a tilt of phase-fronts of the waves propagating in the media [16], and is often discussed in the content of positive vs. negative refractive index materials (see also Refs. [8-12, 22-24]), while its imaginary part defines the wave attenuation or growth. The sign of the square root in Eq. (4) is selected to enforce the causal energy propagation. For dielectrics excited in total internal reflection geometry, as well as for metals and other media with $\mathrm{Re}[k_z^2] < 0$, which do not support propagating waves, the imaginary part of the square root should be always positive regardless of the sign of $\varepsilon$". For other systems, the selection should enforce the wave decay in systems with loss ($\varepsilon$">0) and the wave growth in materials with gain ($\varepsilon$"<0) [23]. This selection of the sign can be achieved by the cut of the complex plane along the negative imaginary axis.

Although such cut of the complex plane is different from the commonly accepted cuts along the positive [22] or negative [8-12,23] real axes, our simulations (Fig.1b) show that this is the only solution guaranteeing the continuity of measurable parameters (such as reflectivity) under the transition from a weak loss regime to a weak gain regime. Our sign selection is the only one consistent with previous results on gain-assisted reflection enhancement, predictions of gain-assisted SPP behavior [13-15,25], and the experimental data presented below. The implications of selecting different signs of $k_{z2}$ are shown in Fig.1b.

Note that active media excited above the angle of total internal reflection, as well as the materials with $\varepsilon'<0$ and $\varepsilon''<0$ formally fall under negative index materials category. However, since $|k_z''|$ in this case is greater than $|k_z'|$, the "left handed" wave experiences very large attenuation (in the presence of gain!!!), which in contrast to claims of Ref. [22], makes the material unsuitable for superlenses and other proposed applications of NIMs [8-12].

In the limit of small plasmonic loss/gain, when the decay length of SPP, $L$, is much greater than $2\pi/k_x^{0'}$, and in the vicinity of $\theta_0$, Eq. (3) can be simplified, revealing the physics behind the gain-assisted plasmonic loss compensation:

$$R(\theta) \approx |r_{01}|^2 \left[ 1 - \frac{4\gamma_i\gamma_r + \delta(\theta)}{(k_x - k_x^0 - \Delta k_x^0)^2 + (\gamma_i + \gamma_r)^2} \right], \quad (5)$$

where $\xi = \frac{c(\varepsilon_2' - \varepsilon_1')}{2\omega} \left( \frac{\varepsilon_2' + \varepsilon_1'}{\varepsilon_2'\varepsilon_1'} \right)^{3/2}$, $r_{01}^0 = r_{01}(\theta_0)$, and $\delta(\theta) = 4(k_x - k_x^0 - \Delta k_x^0)\text{Im}(r_0)\text{Im}(e^{i2k_z^0 d_1})/\xi$.

The shape of $R(\theta)$ is dominated by the Lorentzian term in Eq. (5). Its width is determined by the propagation length of SPP,

$$L = \left[ 2(\gamma_i + \gamma_r) \right]^{-1}, \quad (6)$$

which, in turn, is defined by the sum of the internal (or propagation) loss

$$\gamma_i = k_x^{0"} = \frac{\omega}{2c}\left(\frac{\varepsilon_1'\varepsilon_2'}{\varepsilon_1' + \varepsilon_2'}\right)^{3/2}\left(\frac{\varepsilon_1''}{\varepsilon_1'^2} + \frac{\varepsilon_2''}{\varepsilon_2'^2}\right). \quad (7)$$

and the radiation loss caused by SPP leakage into the prism,

$$\gamma_r = \mathrm{Im}\left(r_{01}e^{i2k_z^0 d_1}\right)/\xi. \quad (8)$$

The radiation loss also leads to the shift of the extremum of the Lorentzian profile from its resonant position $k_x^0$,

$$\Delta k_x^0 = \mathrm{Re}\left(r_{01}e^{i2k_z^0 d_1}\right)/\xi. \quad (9)$$

The term $\delta$ in Eq. (5) results in the asymmetry of $R(\theta)$.

The excellent agreement between exact Eq. (3), solutions of Maxwell equations using transfer matrix method [26] and approximate Eqs. (5,6) for the 60 nm silver film are shown in Figs. 1,3.

The gain in the medium reduces internal loss $\gamma_i$ of SPP, Eq. 7. In reasonably thick metallic films (where $\gamma_i > \gamma_r$ in the absence of gain) the "dip" in the reflectivity profile $R_{min}$ is reduced when gain is first added to the system, reaching $R_{min}=0$ at $\gamma_i \approx \gamma_r$ (Fig. 2a). With further increase of gain, $\gamma_i$ becomes smaller than $\gamma_p$, leading to an increase of $R_{min}$. The resonant value of $R$ is equal to unity when *internal* loss is completely compensated by gain ($\gamma_i=0$) at

$$\varepsilon_2'' = -\frac{\varepsilon_1''\varepsilon_2'^2}{\varepsilon_1'^2}. \quad (10)$$

In the vicinity of $\gamma_i=0$, the reflectivity profile is dominated by the asymmetric term $\delta$. When gain is increased to even higher values, $\gamma_i$ becomes negative and the dip in the reflectivity profile converts into a peak, consistent with predictions of Refs. [13,14]. The peak has a singularity when the gain compensates *total* SPP loss ($\gamma_i+\gamma_r=0$). Past the singularity point, the system becomes unstable and cannot be described by stationary Eqs. (3-5) [27]. Instead, one should consider the rate equations describing populations of energy states of dye as well as a coupling between excited molecules and the SPP field. In thin metallic films (when $\gamma_i<\gamma_r$ at $\varepsilon_2''=0$), the resonant value of $R$ monotonically grows with the increase of gain, Fig. 2b.

Experimentally, SPPs were studied in the attenuated total internal reflection setup of Fig. 1a. The 90° degree prism was made of glass with the index of refraction $n_0=1.784$. Metallic films were produced by evaporating 99.99% pure silver.

Rhodamine 6G dye (R6G) and polymethyl methacrylate (PMMA) were dissolved in dichloromethane. The solutions were deposited to the surface of silver and dried to a film. In the majority of experiments, the concentration of dye in dry PMMA was equal to 10 g/l ($2.1\times10^{-2}$ M) and the thickness of the polymer film was of the order of 10 μm.

The prism was mounted on a motorized goniometer. The reflectivity $R$ was probed with $p$ polarized He-Ne laser beam at $\lambda=594$ nm. The reflected light was detected by a photodiode or a photomultiplier tube (PMT) connected to the integrating sphere, which was moved during the scan to follow the walk of the beam.

The permittivity of metallic film was determined by fitting the experimental reflectivity profile $R(\theta)$ of not pumped system with Eq.(3), inset of Fig. 4a. As a rule, experimental values $\varepsilon_1'$ and $\varepsilon_1''$ did not coincide with the commonly used data of Ref. [28].

In the measurements with optical gain, the R6G/PMMA film was pumped from the back side of the prism (Fig. 1a) with Q-switched pulses of the frequency doubled Nd:YAG laser ($\lambda=532$ nm, $t_{pulse}=10$ ns, repetition rate 10 Hz). The pumped spot, with the diameter of ~3 mm, completely overlapped the smaller spot of the He-Ne probe beam. Reflected He-Ne laser light was directed to the entrance slit of the monochromator, set at $\lambda=594$ nm, with PMT attached to the monochromator's exit slit. Experimentally, we recorded reflectivity kinetics $R(\theta,t)$ under short-pulsed pumping at different incidence angles (Fig. 4b).

In samples with relatively thin ( 40 nm) metallic films, strong emission signal from the R6G-PMMA film was observed in the absence of He-Ne probe beam. We therefore performed two measurements of kinetics for each data point: one in the absence of the probe beam, and one in the presence of the beam. We then subtracted "emission background" (measured without He-Ne laser) from the combined reflectivity and emission signal. The kinetics measurements had a relatively large data scatter, which was partially due to the instability of the Nd:YAG laser.

The results of the reflectivity measurements in the 39 nm silver film are summarized in Fig. 4a. Two sets of data points correspond to the reflectivity without pumping (measured in flat parts of the kinetics before the laser pulse) and with pumping (measured in the peaks of the kinetics). By dividing the values of $R$ measured in the presence of gain by those without gain, we calculated the relative enhancement of the reflectivity signal to be as high as 280% – a significant improvement in comparison to Ref. [19], where the change of the reflectivity in the presence of gain did not exceed 0.001%. Fitting both reflectivity curves with Eq. (3) and known $\varepsilon'_1$=-15, $\varepsilon''_2$=0.85 and $\varepsilon_2'= n_2^2$=2.25, yields $\varepsilon_2''$ -0.006. For $\lambda$=594 nm, this corresponds to optical gain of $g$=420 cm$^{-1}$.

In thicker silver films, calculations predict initial reduction of the minimal reflectivity $R(\theta)$ at small values of gain followed by its increase (after passing the minimum point $R$=0) at larger gains, Fig. 2a. The predicted reduction of $R$ was experimentally observed in the 90 nm thick film, where instead of a peak in the reflectivity kinetics, we observed a dip, inset of Fig. 4b.

For the silver film parameters measured in our experiment, Eq. (5) predicts complete compensation of *intrinsic* SPP loss at optical gain of 1310 cm$^{-1}$. For a better quality silver characterized by the dielectric constant of Ref. [28], the critical gain is smaller, equal to 600 cm$^{-1}$. In addition, if a solution of R6G in methanol ($n$=1.329) is used instead of the R6G/PMMA film, then the critical value of gain is further reduced to 420 cm$^{-1}$. This is the value of gain achieved in our experiment. Thus, in principle, at the available gain, one can fully compensate the *intrinsic* SPP loss in silver.

For complete compensation of plasmonic loss in the system, one must also compensate radiation losses. The huge gain equal to 4090 cm$^{-1}$ is required to completely compensate attenuation of SPP in the 39 nm thick film used in our experiment. This value is dramatically reduced in thicker metallic films, since radiative loss strongly depends on the film thickness. For relatively thick ( 100 nm) metallic films, the *total* loss is almost identical to the *internal* loss.

In the experiment described above, the concentration of R6G molecules in the PMMA film was equal to 1.26x10$^{19}$ cm$^{-3}$ (2.1x10$^{-2}$M). Using the spectroscopic parameters known for the solution of R6G dye in methanol and neglecting any stimulated emission effects, one can estimate that 18 mJ laser pulses used in the experiment should excite more than 95% of all dye molecules. At the emission cross section equal to 2.7x10$^{-16}$ cm$^2$ at $\lambda$=594 nm, this concentration of excited molecules corresponds to the gain of 3220 cm$^{-1}$. Nearly eight-fold difference between this value and the one obtained in our experiment is probably due to the combined effects of luminescence quenching of R6G due to dimerization of rhodamine 6G molecules occurring at high concentration of dye [29], and amplified spontaneous emission (ASE). While the detailed study of the ASE-induced effects in the R6G/PMMA-silver systems is beyond the scope of this work, we note that at the value of gain equal to 420 cm$^{-1}$ and the diameter of the pumped spot equal to 3 mm, the optical amplification is enormously large. Obviously, these giant values of the amplification and the gain cannot be maintained in a cw regime, and ASE appears to be a detrimental factor controlling the gain in the pulsed regime. Correspondingly, the choice of a more efficient amplifying medium (as was proposed in Ref. [19]) may not help in compensating the SPP loss by gain.

To summarize, in our study of the propagating surface plasmon polariton in the attenuated total reflection setup, we have established the relationship between (*i*) the gain in the dielectric adjacent to the metallic film, (*ii*) the *internal*, *radiative* and *total* losses, (*iii*) the propagation length of the SPP, and (*iv*) the shape of the experimentally measured reflectivity profile *R(θ)*. We have experimentally demonstrated the optical gain in the dielectric (PMMA film with R6G dye) equal to 420 cm$^{-1}$, yielding nearly threefold increase of the resonant value of the reflectivity. In the case of thick low-loss silver film [28] and low index dielectric, the demonstrated value of gain is sufficient for compensation of the total loss hindering the propagation of surface plasmon polariton.

The work was supported by the NSF PREM grant # DMR 0611430, the NSF CREST grant # HRD 0317722, the NSF NCN grant # EEC-0228390, the NASA URC grant # NCC3-1035, and the Petroleum Research Fund. The authors cordially thank Vladimir M. Shalaev for useful discussions.

**Figure captions**

Fig.1 (a) Schematic of SPP excitation in ATR geometry. (b) Reflectivity $R$ as a function of angle $\theta$. Traces – solutions of exact Eq. (3). Dots – solution of approximate Eq. (6). For all data sets: $\varepsilon_1$=-15.584+0.424i, $d_1$=60 nm. Trace 1: dielectric with very small loss, $\varepsilon_2$=2.25+10$^{-5}$i. Traces 2-4: dielectric with very small gain, $\varepsilon_2$=2.25-10$^{-5}$i. Trace 2 and dots: complex cut along negative imaginary axis (correct; nearly overlaps with trace 1; no discontinuity at the transition from small loss to small gain). Trace 3: complex cut along positive real axis (yields incorrect predictions for incident angles below total internal reflection). Trace 4: complex cut along negative real axis (yields incorrect predictions for incident angles above total internal reflection).

Fig.2. Reflectivity $R$ [Eq.(5)] of the three-layer system depicted in Fig. 1a as a function of angle $\theta$ and pumping (given by imaginary part of $\varepsilon_2$); panel (a) illustrates the evolution of reflectivity in a relatively thick metallic film ($d_1$=70nm); panel (b) corresponds to a thin film ($d_2$=39 nm).

Fig. 3. Inverse propagation length of SPP, $L^{-1}$, in the system depicted in Fig.1a as a function of gain in dielectric, $\varepsilon_2''$. Solid line – solution of Eq. (11), dots – exact numerical solution of Maxwell equations. Top inset: intensity distribution across the system. Bottom inset: Exponential decay of the SPP wave intensity $|E|^2$ (shown in the top inset) along the propagation in the $x$ direction.

Fig. 4. (a) Reflectivity $R(\theta)$ measured without (diamonds) and with (circles) optical pumping in the glass-silver-R6G/PMMA system. Dashed lines – guides for eye. Solid lines – fitting with Eq. (3) at $\varepsilon_0'=n_0^2$=1.784$^2$=3.183, $\varepsilon_0''$=0, $\varepsilon_1'$=-15, $\varepsilon_1''$=0.85, $d_1$=39 nm, $\varepsilon_2'= n_2^2$=1.5$^2$=2.25, $\varepsilon_2''$ 0 (trace 1) and $\varepsilon_2''$ -0.006 (trace 2). Inset: Angular reflectivity profile $R(\theta)$ recorded in the same system without pumping (dots) and its fitting with Eq. (3) (solid line). (b) Reflectivity

kinetics recorded in the glass-silver-R6G/PMMA structure under pumping. The angle $\theta$ corresponds to the minimum of the reflectivity; $d_1$=39 nm. Inset: Reflectivity kinetics recorded in a thick film ($d_1$=90 nm) shows a 'dip' at small values of gain.

Fig. 1

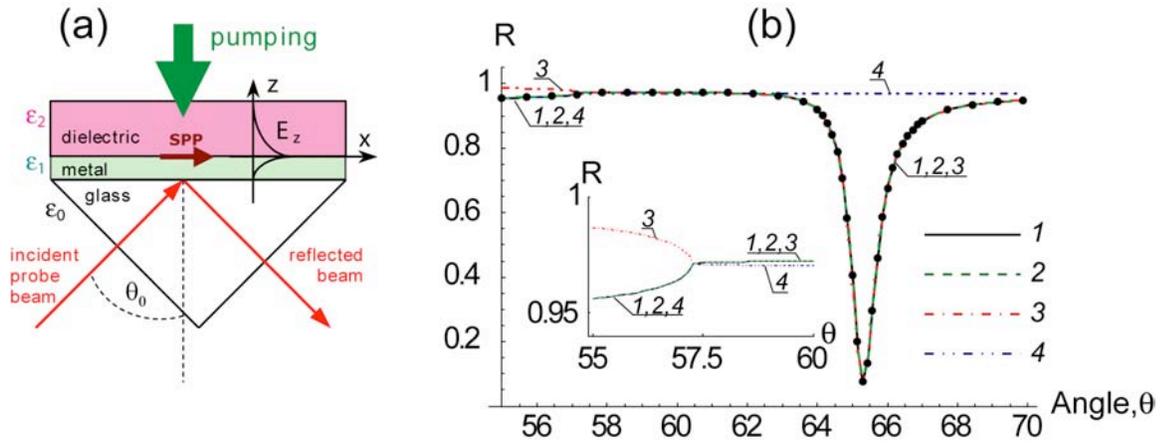

Fig.1 (a) Schematic of SPP excitation in ATR geometry. (b) Reflectivity *R* as a function of angle *θ*. Traces – solutions of exact Eq. (3). Dots – solution of approximate Eq. (6). For all data sets: $\varepsilon_1$=-15.584+0.424i, $d_1$=60 nm. Trace 1: dielectric with very small loss, $\varepsilon_2$=2.25+10$^{-5}$i. Traces 2-4: dielectric with very small gain, $\varepsilon_2$=2.25-10$^{-5}$i. Trace 2 and dots: complex cut along negative imaginary axis (correct; nearly overlaps with trace 1; no discontinuity at the transition from small loss to small gain). Trace 3: complex cut along positive real axis (yields incorrect predictions for incident angles below total internal reflection). Trace 4: complex cut along negative real axis (yields incorrect predictions for incident angles above total internal reflection).

Fig. 2.

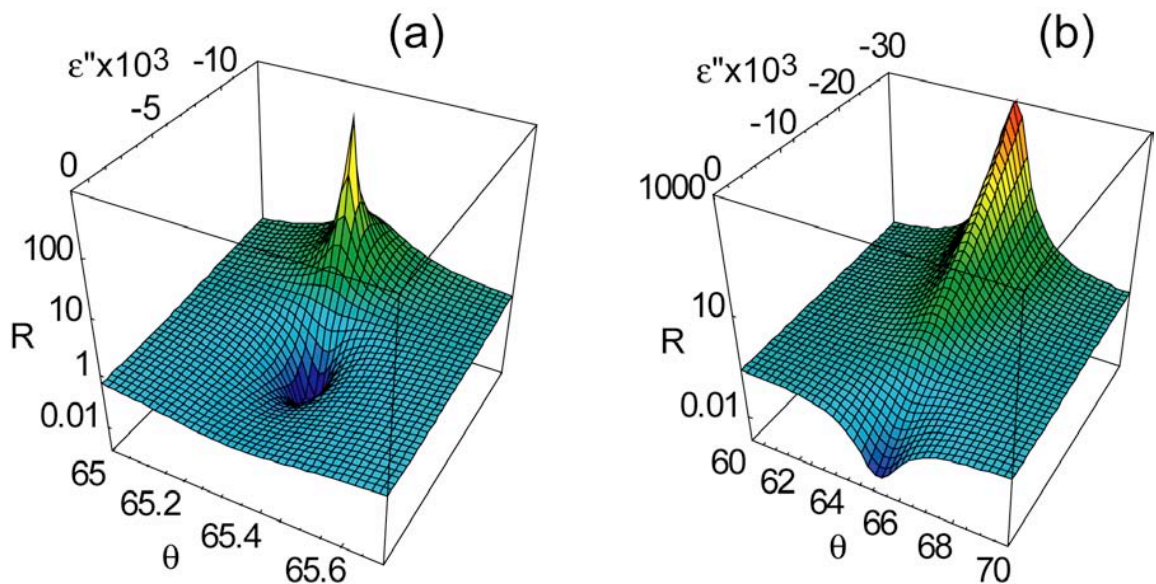

Fig.2. Reflectivity $R$ [Eq.(5)] of the three-layer system depicted in Fig. 1a as a function of angle $\theta$ and pumping (given by imaginary part of $\varepsilon_2$); panel (a) illustrates the evolution of reflectivity in a relatively thick metallic film ($d_1$=70nm); panel (b) corresponds to a thin film ($d_2$=39 nm).

Fig. 3.

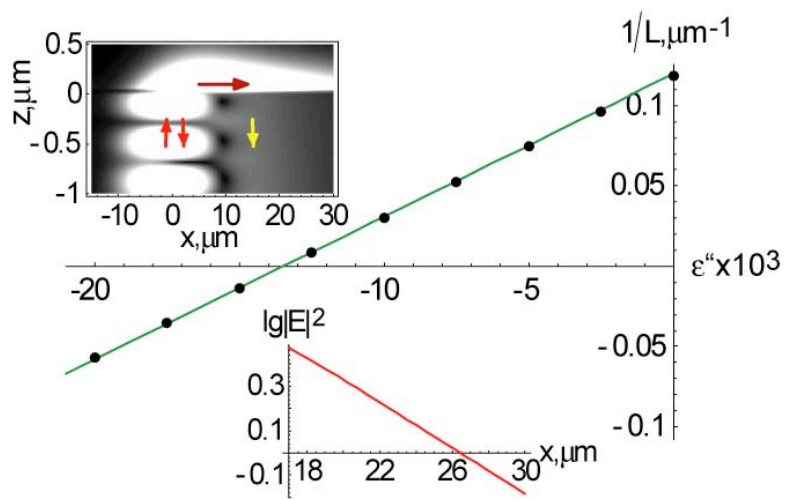

Fig. 3. Inverse propagation length of SPP, $L^{-1}$, in the system depicted in Fig.1a as a function of gain in dielectric, $\varepsilon_2''$. Solid line – solution of Eq. (11), dots – exact numerical solution of Maxwell equations. Top inset: intensity distribution across the system. Bottom inset: Exponential decay of the SPP wave intensity $|E|^2$ (shown in the top inset) along the propagation in the $x$ direction.

Fig. 4.

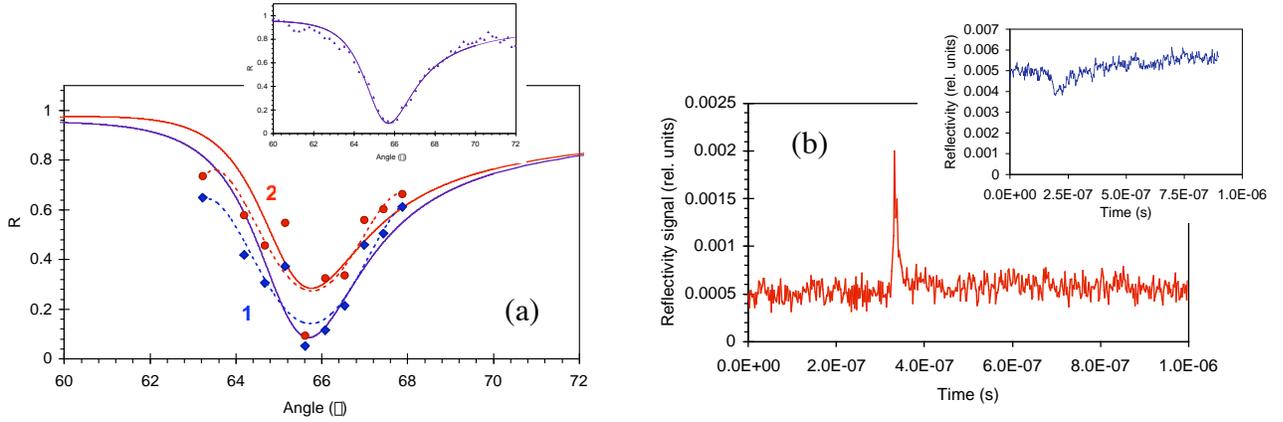

Fig. 4. (a) Reflectivity $R(\theta)$ measured without (diamonds) and with (circles) optical pumping in the glass-silver-R6G/PMMA system. Dashed lines – guides for eye. Solid lines – fitting with Eq. (3) at $\varepsilon_0'=n_0^2=1.784^2=3.183$, $\varepsilon_0''=0$, $\varepsilon_1'=-15$, $\varepsilon_1''=0.85$, $d_1=39$ nm, $\varepsilon_2'=n_2^2=1.5^2=2.25$, $\varepsilon_2''$ 0 (trace 1) and $\varepsilon_2''$ -0.006 (trace 2). Inset: Angular reflectivity profile $R(\theta)$ recorded in the same system without pumping (dots) and its fitting with Eq. (3) (solid line). (b) Reflectivity kinetics recorded in the glass-silver-R6G/PMMA structure under pumping. The angle $\theta$ corresponds to the minimum of the reflectivity; $d_1=39$ nm. Inset: Reflectivity kinetics recorded in a thick film ($d_1=90$ nm) shows a 'dip' at small values of gain.